\documentclass[prb,twocolumn,showpacs,showkeys,amsmath,amssymb,floatfix]{revtex4}
\usepackage{dcolumn,graphicx}
\usepackage{bm}
\def\corund{$\alpha$-Al$_2$O$_3$}
\def\ruby{$\alpha$-Al$_2$O$_3$:Cr$^{3+}$}
\def\pinksapph{$\alpha$-Al$_2$O$_3$:Ti$^{3+}$}
\def\bluesapph{$\alpha$-Al$_2$O$_3$:Fe-Ti}
\def\tio{$\alpha$-Ti$_2$O$_3$}
\def\esko{$\alpha$-Cr$_2$O$_3$}
\def\hemat{$\alpha$-Fe$_2$O$_3$}
\def\rutile{TiO$_2$}

\begin{document}

\title{Structural relaxations around Ti, Cr and Fe impurities in 
$\alpha$-Al$_2$O$_3$ probed by x-ray absorption near edge 
structure combined with first-principles calculations}

\author{Emilie Gaudry, Delphine Cabaret, Philippe Sainctavit, 
        Christian Brouder, Francesco Mauri,}
\affiliation{ Laboratoire de Min\'eralogie-Cristallographie\\
               UMR CNRS 7590, Universit\'e Pierre et Marie Curie, 
               case 115, 4 place Jussieu\\
               F-75252 Paris Cedex 05, France}
\author{Jos\'e Goulon, Andrei Rogalev}
\affiliation{ European Synchrotron Radiation Facility\\
              BP 220, F-38043 Grenoble Cedex, France}

\date{\today}

\begin{abstract}
We determine the structural relaxations around
paramagnetic impurities (Ti, Cr, Fe) in corundum 
($\alpha$-Al$_2$O$_3$), by combining x-ray absorption near edge 
structure (XANES) experiments and \textit{ab initio}
calculations.
The structural relaxations are found to be very local.
We then show that XANES is sensitive to small
variations in interatomic distances  
within the coordination shell of the absorbing atom.
The experiments were carried out on single crystals of ruby and sapphires.
Linear dichroic signals are essential to characterize
the geometry of the impurity site.
The calculations were performed within
a self-consistent ``non muffin-tin'' framework, that uses
pseudopotentials, plane-wave basis set, and 
the continued fraction for the absorption cross section.
\end{abstract}

                            
\maketitle

\section{\label{sec:intro}Introduction}

The presence of impurities in a crystal can influence its
mechanical, magnetic, semiconducting, superconducting,
dielectric, spectroscopic and transport properties.
To cite just a few specific examples, impurities can
improve the voltage holdoff capability of
insulating materials\cite{Miller81}, they
are critical for the optical properties of
most laser materials and gemstones, and they
turned out to be the secret of ancient Damascus steel
blades.\cite{Verhoeven}

Still, the physics of impurity systems is not well developed.
The local crystallographic structure around impurities
is unknown in most cases, although it is an essential piece of
information to undestand their influence on the physical
properties of the host and to carry out \textit{ab initio}
simulations of these materials. This situation is due
to the experimental and theoretical difficulties that
are met to obtain microscopic information on impurity
systems. From the theoretical point of view, quantum
calculations of impurity systems require the use
of large supercells that have long not been computationally
manageable. From the experimental point
of view, impurity systems can only be measured with
methods which can selectively probe certain atomic
species. Among such methods, many of them, such as
electronic paramagnetic resonance, give only indirect
information on the position and nature of the atoms
surrounding the impurity. Due to the advent of 
third-generation synchrotron radiation facilities,
x-ray absorption spectroscopy (XAS) is now able to
investigate impurities in solids.

The extended x-ray absorption fine structure (EXAFS) region of 
a XAS spectrum provides quantitative information about the
short range organization around the absorbing atom (coordination number,
interatomic distances). The x-ray absorption near-edge structure
(XANES) region usually gives
qualitative information about the atomic arrangement up to the medium
range order. XANES is sensitive to the electronic structure (bonding)
and probes the empty states of solids.
In the case of dilute elements in single crystals, 
especially at the $K$ edge of 
3$d$ transition elements, collecting 
EXAFS spectra with a good signal-to-noise ratio is a difficult
task, because the intensity of diffraction peaks becomes 
much larger than the EXAFS signal.
Taking the example of impurities in corundum \corund,\cite{Gaudry03}
good EXAFS spectra could be collected for 10000 wt.ppm Cr$^{3+}$
impurities, but the usable energy range obtained for
1500 wt.ppm Fe$^{3+}$ impurities was already quite limited,
and the concentration of coloring impurities is often
an order of magnitude lower than this.
In such cases, XANES becomes 
a practicle and precious technique to get information 
about the local structure of the absorbing atom.\cite{Waychunas03} 
However, the analyzis of the experimental XANES
data is not straightforward because of 
the photoelectron multiple-scattering processes that occur
in the near-edge region. \textit{Ab initio}
XANES simulations are then required
to relate the experimental spectral features to
the local geometry around the aborbing atom.

Recently, an efficient first-principle approach based on plane-wave 
pseudopotential formalism has been developed to 
calculate $K$-edge XANES spectra.\cite{Taillefumier02}
Here we apply this method to 
the $K$-edge of substitutional paramagnetic impurities 
in aluminum oxide. 
More precisely, we investigate the structural modifications
of the corundum crystal structure 
induced by the presence of substitutional Ti$^{3+}$, 
Cr$^{3+}$ or Fe$^{3+}$. This system is chosen because
corundum containing transition metal impurities 
is important for laser applications and as gemstones.
Colorless corundum (\corund)
becomes red ruby, pink sapphire, or yellow sapphire 
when a small amount of Cr$^{3+}$, Ti$^{3+}$, or Fe$^{3+}$  
ions substitute for
Al$^{3+}$ ions, respectively.\cite{Burns93, Nassau83} 
The color of blue sapphire is due to the presence of
(Fe-Ti) pairs in \corund. 

For this investigation, polarized XANES spectra are measured
and calculated. Corundum belongs to the trigonal Bravais lattice, 
the optical axis being parallel
to the [111] direction of the trigonal unit cell
(i.e. to the $c$-axis of the hexagonal cell).
Corundum is then a dichroic compound in the electric dipole 
approximation.
The dichroic signal is a direct signature of the departure
of the impurity site from octahedral symmetry.\cite{Gaudry03}
XANES calculations are performed from structural models
resulting from \textit{ab initio} energy minimization calculations.
The agreement of our experimental and theoretical spectra
demonstrates that precise structural information can be extracted
from the anglular-dependent XANES spectra, provided 
the electronic potential is accurately modelled.

\section{\label{sec:experiments}Experiments}

\subsection{\label{subsec:samples}Crystallographic structure of \corund\ 
and samples description}

\begin{figure}[!ht]
\includegraphics[scale=0.5]{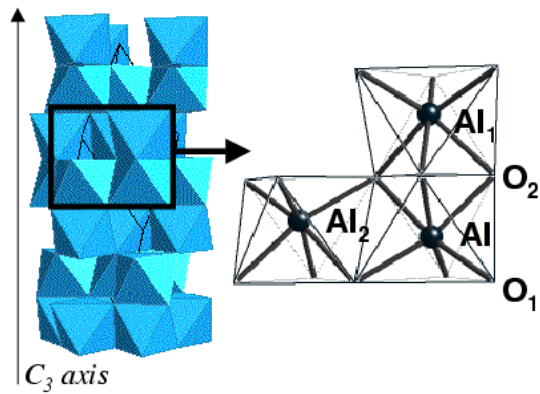}
\caption{\label{corund} Aluminum atomic site in corundum. Left:
the corundum structure is represented as a stacking of AlO$_6$
octahedra (the trigonal cell and the three-fold symmetry axis
are indicated). Right: zoom of the black-framed region of
the left panel. The aluminum site with its O$_1$, O$_2$,
Al$_1$ and Al$_2$ first four neighbors are indicated.}
\end{figure}

Corundum (\corund) belongs to the $R\bar{3}c$
($D_{3d}^6$) space group.\cite{Newnham62} The trigonal unit cell
contains two Al$_2$O$_3$ formula units.
The atomic environment of aluminum is illustrated in Fig.~\ref{corund}.
The atomic site of aluminum is a distorded octahedron with a 3 ($C_3$)
local point symmetry. The AlO$_6$ octahedron is characterized
by two different Al-O interatomic distances.
If we consider that the absorbing atom is denoted by Al
in Fig.~\ref{corund}, the nearest three oxygen
atoms are labeled O$_1$ in Fig.~\ref{corund}.
The farther three oxygen atoms, labeled O$_2$,
form a face shared by two octahedra along
the three-fold symmetry axis. Beyond the coordination shell
of aluminum, the next two neighbors are aluminum atoms, labeled
Al$_1$ and Al$_2$ in Fig.~\ref{corund}, and are relative to
face-shared octahedra and edge-shared octahedra, respectively.

Three synthetic single crystals of doped \corund\ (obtained
by the Verneuil process) are used for this study: a red ruby 
(\ruby), a blue  sapphire (\bluesapph) and a pink sapphire (\pinksapph). 
The impurity concentration in each sample is given in table~\ref{tab:Compo}.
They are measured with the Cameca Microbeam electron
microprobe at the CAMPARIS analytical facility of Paris. A 30~kV
acceleration with a 15~nA beam current,
defocused to 10$\mu$m, is used. X-ray intensities are corrected for
dead-time, background, and matrix effects using the Cameca ZAF routine.
The standards used are \corund, \esko, \hemat\ and \rutile.
The blue sapphire sample is chosen to represent the model compound
of iron impurity in corundum (\corund:Fe$^{3+}$). 
It is indeed preferred to a yellow 
sapphire sample, that only contains iron impurities, because 
synthetic yellow sapphires are known to be inhomogeneous.\cite{Eigenmann70} 
Consequently, in this study, we assume that the low concentration of 
titanium (750 wt.ppm) compared to that of iron (1500 wt.ppm) does not 
affect the site relaxation around iron.

\begin{table}[!ht]
\caption{\label{tab:Compo}Impurity concentration (in wt.ppm) in the three
doped \corund \ samples.}
\begin{ruledtabular}
\begin{tabular}{lccc}
Samples       & Cr (wt.ppm)       & Fe (wt.ppm)     & Ti (wt.ppm)     \\
\hline
Red ruby      & 10000 ($\pm$ 1000)&    -            &    -            \\
Blue sapphire &    -              & 1500 ($\pm$ 50) &  750 ($\pm$ 30) \\
Pink sapphire &   -               &   -             &  540 ($\pm$ 30) \\
\end{tabular}
\end{ruledtabular}
\end{table}

The samples are cylindrically shaped: 15~mm diameter and 3~mm thickness 
for the ruby sample, and 5~mm diameter and 1~mm thickness for the sapphire
samples. They are cut so that the optical axis 
is in the plane of the disk surface, and the [10$\bar{1}$]
direction of the trigonal cell is orthogonal to the 
disk surface.

\subsection{\label{subsec:xanes} X-ray absorption measurements}

X-ray absorption measurements are carried out at the European Synchrotron 
Radiation Facility (ESRF) on the ID12 beam line, which is dedicated to 
polarization dependent spectroscopies.\cite{Goulon98,Rogalev01} 
The monochromatic x-ray beam is obtained through a double Si(111)
crystal monochromator, with a resolution $\Delta E/E \simeq 10^{-4}$.
The oriented samples are placed on a rotating holder with the normal 
of the disk surface parallel to the x-ray wave vector and
to the rotation axis.
The fluorescence intensity is measured by eight detectors symmetrically 
set around the x-ray beam. 

XANES spectra are recorded for several angles $\theta$ between the 
optical axis of the sample and the x-ray polarization vector. We note 
$\sigma_{\parallel}$ and $\sigma_{\perp}$ the absorption cross section
obtained with $\theta=0^\circ$ and $\theta=90^\circ$, respectively.
In the electric dipole 
approximation, the isotropic signal is given by 
$(\sigma_{\parallel}+2\sigma_{\perp})/3$, and the 
dichroic signal is given by the difference
$\sigma_{\parallel}-\sigma_{\perp}$. 
The measurement of x-ray absorption spectra of impurities
in crystals is impeded by the presence of large diffraction
peaks.
In order to decrease the intensity of diffraction peaks and
elastic scattering, V and Mn filters (50~$\mu$m width) are
used for the Cr $K$ edge in ruby and the Fe $K$ edge in blue sapphire,
respectively.
By so doing, the spectra
recorded for the two angles $\theta=0^\circ$ and $\theta=90^\circ$
are not affected by diffraction peaks.
On the contrary, no filter is available for the Ti $K$ edge
and the absorption was measured for
100 angles $\theta$ from $0$ to $360^\circ$. The $\sigma_\parallel$ and 
$\sigma_\perp$ spectra are reconstructed from this amount of data 
by using a Fourier decomposition to remove 
the contribution of diffraction peaks to the spectra.

\subsection{Calculation methods}

The starting point of any XANES \textit{ab initio} calculation is a
structural model. Two structural 
models are considered for each compound. 
The first one, called \textit{non-relaxed} model,
corresponds to the substitution of one aluminum atom 
by one impurity Ti, Cr or Fe in the \corund,
without relaxing its structure.
The second one, 
called \textit{relaxed} model, is the result of {\it ab initio} 
energy minimization calculation carried out on the 
\textit{non-relaxed} model. This calculation is performed 
using Car-Parrinello 
molecular dynamics with simulated annealing\cite{Car85} as 
implemented in the \texttt{CPMD} code.\cite{cpmd} 
The XANES calculations are then carried out within the scheme
of the \texttt{PARATEC} code,\cite{paratec} using the formalism 
described in Ref.\cite{Taillefumier02} Both codes are 
based on the density functional theory (DFT) within the spin 
polarized local density approximation (LSDA), and use plane-wave 
basis set and norm-conserving Troullier-Martins 
pseudopotentials\cite{Troullier91} in the Kleinman-Bylander 
form.\cite{Kleinman82} 
In the following, we give all the parameters used in both kinds
of calculation.

\subsubsection{\label{subsec:StrucModel} Structural model construction}

The structural models are constructed from the lattice 
parameters and atomic positions resulting from a previous 
\textit{ab initio} calculation, performed 
by Duan \textit{et al.}\cite{Duan98a,Duan98b} in \corund.
In Ref.\cite{Duan98a},
the trigonal unit cell parameters are $a_R=5.11$~\AA\ and
$\alpha= 55.41^\circ$, the aluminum atoms are
in 4~$c$ symmetry sites\cite{crystallo}
 with $x=0.352$ and the oxygen atoms are in
6~$e$ symmetry sites\cite{crystallo} with $x=0.555$.
We verify by using \texttt{CPMD} that this corundum optimized structure 
obtained by Duan \textit{et al.} corresponds to atomic forces 
less than $2.10^{-3}$~Ry/\AA.  
Since we use periodic boundary conditions, we consider
$2\times2\times2$ supercells. The
supercells contain 80 atoms: 48 oxygen atoms, 31 aluminum
atoms and 1 transition metal atom (Ti, Cr or Fe)
in substitution for aluminum.
With such large supercells, the interaction between two impurities
(belonging to two neighboring cells) is negligible.
These supercells define the \textit{non-relaxed} models
mentioned above.
The \textit{relaxed} models are obtained by 
minimization of the energy of the \textit{non-relaxed} supercells.
All atoms are allowed to relax, while the lattice constants are
fixed. 
The impurities considered in this study are known to be
in a high-spin state,\cite{Lever} therefore
the spin multiplet imposed on the trivalent
impurities are S=$\frac{1}{2}$ for
Ti$^{3+}$ ([Ar]$4s^03d^1$), S=$\frac{3}{2}$ for Cr$^{3+}$ ([Ar]$4s^03d^3$),
and S=$\frac{5}{2}$ for Fe$^{3+}$ ([Ar]$4s^03d^5$).
The parametrization of the norm-conserving
Troulliers-Martins pseudopotentials is given in Table~\ref{Pseudo}.
The wave functions and the charge density were expanded in plane waves
with cutoff of 80~Ry and 320~Ry, respectively. Since the supercell 
is rather large, and since the systems are insulating materials,
the Brillouin zone is only sampled at the $\Gamma$ point.  

\begin{table}[!ht]
\caption{\label{Pseudo} Parametrization used for the generation of the
pseudopotentials. The core radii of
the valence states are indicated between parenthesis in \AA.}
\begin{ruledtabular}
\begin{tabular}{cccccc}
atom     & Al         & O          & Ti         & Cr         & Fe         \\
\hline
valence & $3s$ (1.06)& $2s$ (0.77)& $3s$ (0.58) & $3s$ (0.53) & $3s$ (0.48)\\
states  & $3p$ (1.06)& $2p$ (0.77)& $3p$ (0.90) & $3p$ (0.79) & $3p$ (0.90)\\
        & $3d$ (1.06)&            & $3d$ (0.90) & $3d$ (0.79) & $3d$ (0.90)\\
\hline
local part& $d$      & $p$        &  $d$  &  $d$  & $d$        \\
\end{tabular}
\end{ruledtabular}
\end{table}

\subsubsection{\label{subsec:XANESCalc} XANES calculations}

The method used for XANES calculations has been already described in
Ref.\cite{Taillefumier02,Cabaret04} Therefore its main aspects 
are recalled here. The method uses periodic boundary conditions,
plane-wave basis-set, pseudopotentials and reconstructs all-electron 
wave functions within the projector augmented wave (PAW) 
framework.\cite{Blochl94} 
In order to allow the treatment of large supercells
(hundreds of atoms), the scheme uses a recursion method to 
construct a Lanczos basis and then compute the cross section as a 
continued fraction,\cite{Haydock72,Haydock75} in the electric dipole
and quadrupole approximations.\cite{E1-E2} Electric quadrupole
transitions are relevant for the pre-edge region, in particular
at the $K$-edge of transition metals.
The absorption cross section
is calculated beyond the ``muffin-tin'' approximation, that is known
to limit the applications of the multiple scattering theory 
traditionally used for XANES simulations.\cite{Feff8,Continuum,MXAN}
\begin{figure}[!ht]
\includegraphics{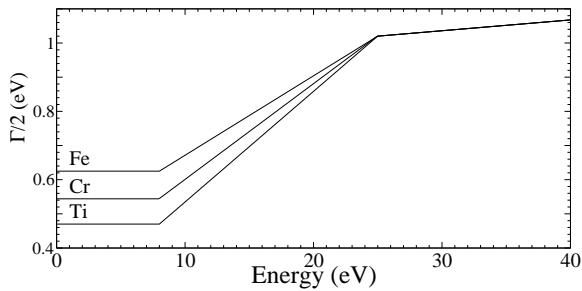}
\caption{\label{Gamma-fig}
Energy dependence of the $\gamma$ parameters
used for the convolution of the XANES spectra at the Ti, Cr and Fe
$K$ edges in doped \corund. The zero in the energy scale refers to the
highest occupied level in the corresponding structure.}
\end{figure}

XANES spectra are computed at the transition metal $K$ edge for both
\textit{relaxed} and \textit{non-relaxed} models of the three compounds,
using the 80-atom supercells described in the previous subsection.  For the
pseudopotentials construction, we again use the parametrization given in
Table~\ref{Pseudo}.  In order to take into account the core-hole effects in
the calculations, the Ti, Cr and Fe pseudopotentials are generated with only
one $1s$ electron. The spin multiplet degeneracy is set by imposing the
number of up and down states given by the \texttt{CPMD} code.  Convergence of
the XANES theoretical spectra is reached for the following set of parameters:
a 64~Ry energy cutoff and one ${\bm k}$-point for the self-consistent
spin-polarized charge density calculation, and 32 ${\bm
k}$-points\cite{Monkhorst76} for the absorption cross section calculation.
An energy dependent broadening parameter $\gamma$ is used in the continued
fraction (see Ref.~\cite{Taillefumier02,Cabaret04}) 
in order to account for the two main photoelectron damping
modes.\cite{Kokubun04} This width $\gamma$ corresponds through the
time-energy uncertainty relations to a lifetime of the photoelectron.  At low
energy, the lifetime of the photoelectron is only limited by the core-hole
lifetime within the plasmon pole approximation. The parameter $\gamma$ was
then set to a constant, which corresponds to $K$ level full width at half
maximum  given by Krause and Oliver.\cite{Krause79} At higher energy, the
kinetic energy of the photoelectron is great enough so that it can interact
with the electron gas of the system. Its amplitude is damped, provoking a
sharp decrease in the lifetime. Therefore the  $\gamma$ parameter was set to
rapidly increase, and then to have a smooth linear behaviour. The $\gamma$
energy-dependent parameters used in this study are displayed in
Fig.~\ref{Gamma-fig}. 

\section{\label{sec:ResultDisc} Results and discussion}

This section is organized as follows. First, we present the results
of the structural relaxation around the substitutional Ti$^{3+}$,
Cr$^{3+}$ and Fe$^{3+}$ ions in \corund\ (Sec.\ref{subsec:res_models}).
Second, we discuss the influence of
the core-hole effects in the case of the
Cr $K$-edge in ruby (Sec.\ref{subsec:pot}).
Third, we confront with experiments the calculated spectra obtained
with both \textit{relaxed} and \textit{non-relaxed} models
(Sec.\ref{subsec:exp_calc}). 

\subsection{\label{subsec:res_models} Structural model analysis}

The \textit{relaxed} models of \ruby\ and \corund:Fe$^{3+}$ 
are already described and discussed in a previous 
work.\cite{Gaudry03} However, in order to compare with the  
\textit{relaxed} model of \pinksapph, most
of the results of Ref.\cite{Gaudry03} are recalled here.

The analysis of the relaxation can only be done by careful
comparison of the \textit{relaxed} and \textit{non-relaxed} 
supercells. This is achieved by comparing clusters
that are built from the supercells using the method detailed in 
Ref.\cite{Gaudry03} These clusters are centered on the impurity
and contain 66 atoms (5.2~\AA\ radius). In order to 
avoid the influence of the
impurity displacement in the \textit{relaxed} model, we define 
the mass center $\Omega$ of each cluster by calculating
$\vec{O\Omega}=\frac{1}{\sum_i m_i}\sum_i m_i \vec{OM_i}$, where $O$ is
the impurity position, $M_i$
refers to all the atoms within a given cluster except the impurity
and $m_i$ is the mass of the atom $i$.
For each atom $i$ around the impurity, we evaluate
the norm of the displacement vector, $\vec{V_i}$,
\begin{eqnarray}
|\vec{V_i}|^2 &=&  (X^i_{\mathrm{\textit{relaxed}}}
                   -X^i_{\mathrm{\textit{non-relaxed}}})^2\nonumber \\
              &+&  (Y^i_{\mathrm{\textit{relaxed}}}-
                    Y^i_{\mathrm{\textit{non-relaxed}}})^2\nonumber \\
              &+&  (Z^i_{\mathrm{\textit{relaxed}}}-
                    Z^i_{\mathrm{\textit{non-relaxed}}})^2, \nonumber
\end{eqnarray}  
where ($X^i_{\mathrm{\textit{non-relaxed}}},
        Y^i_{\mathrm{\textit{non-relaxed}}},
        Z^i_{\mathrm{\textit{non-relaxed}}}$)
and ($X^i_{\mathrm{\textit{relaxed}}},
      Y^i_{\mathrm{\textit{relaxed}}},
      Z^i_{\mathrm{\textit{relaxed}}}$) 
are the cartesian coordinates of $\vec{\Omega M_i}$ vector
in the \textit{non-relaxed} and \textit{relaxed} clusters, respectively.
The angular relaxation was also determined. It was defined
by the quantity
$\delta\theta=\theta_{\mathrm{\textit{non-relaxed}}}-
\theta_{\mathrm{\textit{relaxed}}}$, where 
$\theta_{\mathrm{\textit{relaxed}}}$ and
$\theta_{\mathrm{\textit{non-relaxed}}}$
are the angles between the $\vec{\Omega M_i}$ directions 
and the $C_3$ axis in the \textit{relaxed} and \textit{non-relaxed}
clusters, respectively. 
The comparison of the \textit{relaxed} and \textit{non-relaxed} 
clusters leads to the following conclusions.
\begin{figure}[!h]
\includegraphics{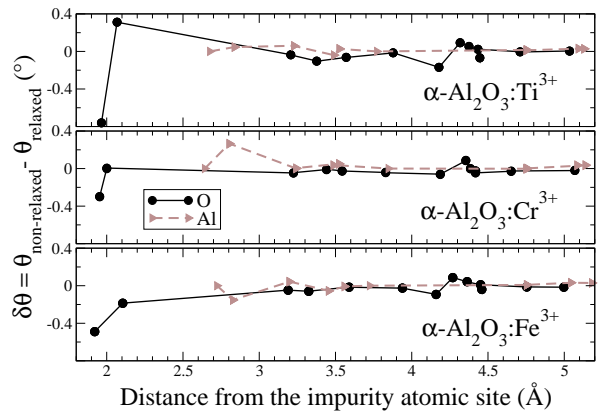}
\caption{\label{fig:angles} Angular relaxation
around the paramagnetic impurities in doped \corund:
$\delta\theta$ angle (see text for explanation)
as a function of distance
from the impurity in the \textit{relaxed} model.}
\end{figure}
\begin{table*}[ht!]
\caption{\label{distances}
M-O and M-Al bond lengths in \corund:M$^{3+}$
(M=Ti, Cr, Fe) issued from the energy minimization calculations
presented in Sec.\ref{subsec:res_models}. These bond lengths
are compared with the experimental and calculated
Al-O and Al-Al interatomic distances in \corund\
and to the M-O and M-M distances in $\alpha$-M$_2$O$_3$. It is worth
noting that the distances indicated for \corund\ relative to Ref.
\cite{Duan98b} are the
M-O and M-Al distances in the \textit{non-relaxed} models.}
\begin{ruledtabular}
\begin{tabular}{lrrrr}
Compound         & \multicolumn{4}{c}{Interatomic distances (in \AA)} \\
                 & \multicolumn{2}{c}{cation-oxygen}
                 & \multicolumn{2}{c}{cation-cation}\\
\hline
\corund\ (exp.\cite{Newnham62})  & d$_{\mathrm{Al-O}_1}$ =1.86
                                 & d$_{\mathrm{Al-O}_2}$ =1.97
                                 & d$_{\mathrm{Al-Al}_1}$=2.65
                                 & d$_{\mathrm{Al-Al}_2}$=2.79    \\
\corund\ (calc.\cite{Duan98b})   & d$_{\mathrm{Al-O}_1}$ =1.84
                                 & d$_{\mathrm{Al-O}_2}$ =1.95
                                 & d$_{\mathrm{Al-Al}_1}$=2.62
                                 & d$_{\mathrm{Al-Al}_2}$=2.77    \\
\pinksapph\ (this calc.)         & d$_{\mathrm{Ti-O}_1}$ =1.97
                                 & d$_{\mathrm{Ti-O}_2}$ =2.07
                                 & d$_{\mathrm{Ti-Al}_1}$=2.72
                                 & d$_{\mathrm{Ti-Al}_2}$=2.83    \\
\ruby\ (calc.\cite{Gaudry03})    & d$_{\mathrm{Cr-O}_1}$ =1.95
                                 & d$_{\mathrm{Cr-O}_2}$ =2.00
                                 & d$_{\mathrm{Cr-Al}_1}$=2.64
                                 & d$_{\mathrm{Cr-Al}_2}$=2.80    \\
\corund:Fe$^{3+}$\ (calc.\cite{Gaudry03})
                                 & d$_{\mathrm{Fe-O}_1}$ =1.92
                                 & d$_{\mathrm{Fe-O}_2}$ =2.10
                                 & d$_{\mathrm{Fe-Al}_1}$=2.72
                                 & d$_{\mathrm{Fe-Al}_2}$=2.83    \\
\tio\ (exp.\cite{Newnham62})     & d$_{\mathrm{Ti-O}_1}$ =2.01
                                 & d$_{\mathrm{Ti-O}_2}$ =2.08
                                 & d$_{\mathrm{Ti-Ti}_1}$=2.59
                                 & d$_{\mathrm{Ti-Ti}_2}$=2.99    \\
\esko\ (exp.\cite{Finger80})     & d$_{\mathrm{Cr-O}_1}$ =1.96
                                 & d$_{\mathrm{Cr-O}_2}$ =2.01
                                 & d$_{\mathrm{Cr-Cr}_1}$=2.65
                                 & d$_{\mathrm{Cr-Cr}_2}$=2.88    \\
\hemat\ (exp.\cite{Finger80})    & d$_{\mathrm{Fe-O}_1}$ =1.94
                                 & d$_{\mathrm{Fe-O}_2}$ =2.11
                                 & d$_{\mathrm{Fe-Fe}_1}$=2.90
                                 & d$_{\mathrm{Fe-Fe}_2}$=2.97    \\
\end{tabular}
\end{ruledtabular}
\end{table*}

First, the impurity site symmetry is conserved:
the Ti, Cr and Fe atoms are still in
a $C_3$ point symmetry site, characterized by two kinds of
M-O$_1$ and M-O$_2$ distances (M = Ti, Cr or Fe).
The angular relaxation
is found to be very small (i.e. less than 0.8 degrees).
This is illustrated
in Fig.~\ref{fig:angles}, which gives $\delta\theta$ as a
function of the distance between the impurity and its neigboring atoms
in the \textit{relaxed} clusters.

Second, a displacement
of the impurity from the initial aluminum site is
observed in the three \textit{relaxed} models.
This displacement occurs along the $C_3$ axis, but its
absolute value depends on the impurity.  Chromium is displaced
by 0.03~\AA\ towards Al$_1$ atom, titanium is displaced
by again 0.03~\AA\ but in the opposite direction, and iron
is further moved (0.09~\AA\ towards Al$_1$ atom). The small
displacement of chromium is in agreement with various experimental
studies of the
literature.\cite{Laurance62,Lohr63,Moss64,McCauley72,Kizler96}
The result for
iron is consistent with an analysis of
electronic paramagnetic resonance (EPR) experiments, that gives
a displacement of $0.04\pm 0.02$~\AA.\cite{Zheng98}
\begin{figure}[!h]
\includegraphics{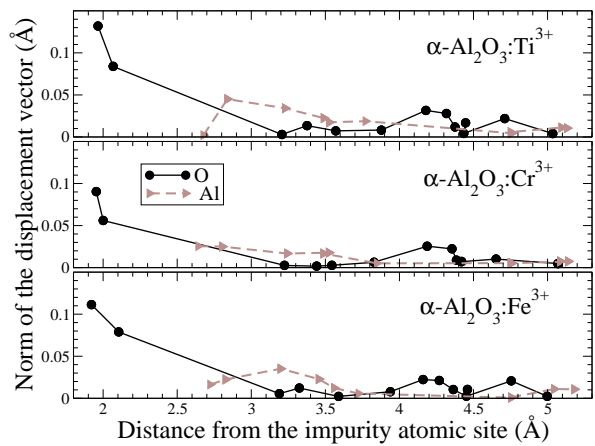}
\caption{\label{fig:dist} Radial relaxation around the
paramagnetic impurities in doped \corund: norm of
atomic displacement vectors
(oxygen atoms in solid line with circles and
aluminum atoms in dashed line with triangles)
as a function of distance
from the impurity in the \textit{relaxed} model.}
\end{figure}

Third, an increase of distances within the
coordination
shell of the impurity is observed. This result is expected
since the ionic radii of Ti, Cr and Fe in octahedral
site\cite{Shannon76}
($r_{\mathrm{Ti}^{3+}}=0.670$~\AA, $r_{\mathrm{Cr}^{3+}}=0.615$~\AA\
and $r_{\mathrm{Fe}^{3+}}=0.645$~\AA) are all greater
than the ionic radius of aluminum ($r_{\mathrm{Al}^{3+}}=0.535$~\AA).
The calculated M-O$_1$ and M-O$_2$
bond lengths in doped \corund\ were found to be close to
M-O$_1$ and M-O$_2$ bond lengths
in the corresponding metal oxides, $\alpha$-M$_2$O$_3$
(\tio, \esko\ and \hemat).
On the contrary, the distances M-Al$_1$ and M-Al$_2$ were less
affected by the relaxation process: they are closer to Al-Al$_1$
and Al-Al$_2$ bond lengths in corundum than the M-M$_1$ and M-M$_2$
distances in the corresponding oxides $\alpha$-M$_2$O$_3$.
Tab.\ref{distances} gathers
the M-O$_1$, M-O$_2$, M-Al$_1$ and M-Al$_2$  distances in
\corund:M$^{3+}$ deduced
from the calculation,
and compares them with
the corresponding distances in \corund, \tio, \esko\ and
\hemat, resulting from x-ray diffraction experiments.
One can remark in Tab.~\ref{distances} that
the Ti-O$_1$ and Ti-O$_2$ bond lengths in \pinksapph\
(1.97~\AA\ and 2.07~\AA) are quite similar to
those in \tio\ (2.01~\AA\ and 2.07~\AA). Besides, the
Ti-Al$_1$ and Ti-Al$_2$ distances in \pinksapph\ (2.72~\AA\ and 2.83~\AA)
are closer to the Al-Al$_1$ and Al-Al$_2$ distances in corundum
(2.65~\AA\ and 2.79~\AA) than to the Ti-Ti$_1$ and Ti-Ti$_2$
distances in \tio\ (2.59~\AA\ and 2.99~\AA).
One can notice that the experimental  distances Cr-Cr$_1$ in \esko\ and
Al-Al$_1$ in \corund\  are both equal to 2.64~\AA. Thus one expects that
Cr-Al$_1$ distance in ruby (\ruby) should be around 2.64~\AA.
The calculated distances of \ruby are slightly larger (+0.015~\AA)
than those determined from EXAFS measurements on powder
samples,\cite{Kizler96} and in good agreement (+0.005~\AA)
with EXAFS measurements performed
and on the same single crystal as the one used in the present
study.\cite{Gaudry03} They are also consistent with data extracted
from optical spectroscopy.\cite{Langer01}
For \corund:Fe$^{3+}$, the calculated distances are
in agreement with EXAFS experiments\cite{Gaudry03} and with
EPR results.\cite{Zheng98}

Finally, the  M-O and M-Al  distances deduced from the
\textit{ab initio} relaxation calculations
suggest that the structural relaxation
around the transition metal impurities is quite local and
mainly concern the coordination shell. This point is confirmed
by the calculations of the atomic displacements
between the \textit{relaxed} and  \textit{non-relaxed}.
Fig.~\ref{fig:dist} represents the norm $|\vec{V_i}|$ as a function of
the distance between the impurity and its neighbors in the \textit{relaxed}
model. It clearly shows that the oxygen coordination shell
absorbs almost the relaxation. The displacement of atoms farther than
2.5\AA\ from the impurity is less than 0.035~\AA.
Comparing the three compounds,
one  also observes  that the relaxation is slightly larger around
titanium and iron than around chromium.

\subsection{\label{subsec:pot} Core-hole effects}

The influence of the core-hole can be described by calculating 
a self-consistent
potential around a core-hole or by solving a Bethe-Salpeter 
equation.\cite{Shirley,Soininen2} For impurity systems,
solving a Bethe-Salpeter equation is probably beyond
the reach of available computer programs. 
Moreover, it was already observed
that both methods give quite similar results.\cite{Taillefumier02}
Therefore, we shall describe the influence of the
core-hole by calculating a self-consistent potential
in the presence of a 1s core-hole.

\begin{figure}[h!]
\includegraphics{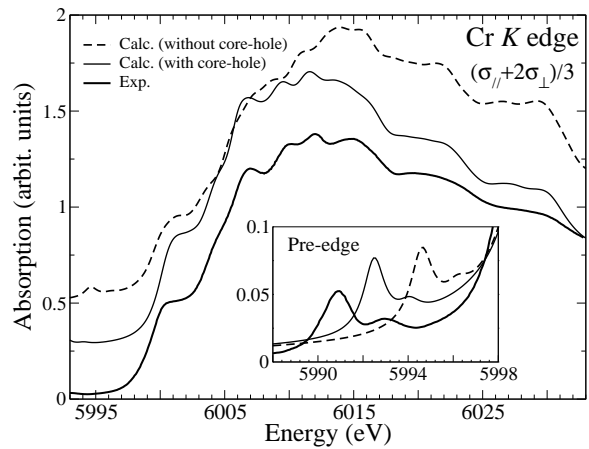}
\caption{\label{trou}
Influence of the core-hole effects in XANES calculation
at the Cr $K$ edge in ruby.
Inset: zoom of the pre-edge region. For pre-edge simulations,
the calculated spectra include both electric dipole and electric
quadrupole transitions (the wave vector $\bm{k}$ is along the [10$\bar{1}$]
direction of the trigonal cell of ruby\cite{quadru}).}
\end{figure}

Calculated spectra performed with or without 
the presence of a $1s$ core-hole on the chromium absorbing atom 
are compared in Fig.~\ref{trou}. It clearly appears that the core-hole 
is needed 
to obtain a good agreement with the experimental curve. In particular, 
the intensity and the energy positions of the features  within
the 6005-6017~eV energy range are not well reproduced
if the core-hole is not taken into account in the calculation. 
We have noticed similar behaviors due to the presence of the core-hole
at the titanium and iron 
$K$-edges in \pinksapph\ and \corund:Fe$^{3+}$, respectively.

On the other hand, the presence of the core-hole has a
non-negligible impact in the pre-edge region.
This is illustrated in the inset of Fig.~\ref{trou}, which represents
the pre-edge region (5985-5998~eV) of the Cr $K$ edge in \ruby.
An interpretation of this
region in terms of group theory has been detailed in Ref.~\cite{Gaudry04}
In the pre-edge region both
electric dipole ($1s\rightarrow p$) and quadrupole
($1s\rightarrow 3d$) transitions occur.
Since the site of the impurity in doped \corund\ is
not centrosymmetric, the $p$ states are present in the pre-edge
through $p-d$ hybridization.
electric dipole ($1s\rightarrow p$) and quadrupole 
($1s\rightarrow 3d$) transitions occur.
Since the site of the impurity in doped \corund\ is
not centrosymmetric, the $p$ states are present in the pre-edge
through $p-d$ hybridization.
The calculated curves shown in the inset of Fig.~\ref{trou}
are the sum of both contributions. 
For the calculation of electric quadrupole transitions, 
the direction of the wave vector of the photon beam, $\bm{k}$,
has to be specified. In the experiments, the single crystals were
placed on the rotating holder so that the wave vector was parallel 
to the [10$\bar{1}$] crystallographic direction. 
Consequently, the spectra $(\sigma_\parallel+2\sigma_\perp)/3$
 shown in Fig.~\ref{trou} are calculated with
$\bm{k}/\!/[10\bar{1}]$.\cite{quadru} The two features
of the experimental pre-edge are reproduced with or without the
1$s$ core-hole (see inset of Fig.~\ref{trou}).
The presence of the core-hole essentially provokes
a 2.5~eV shift towards lower energy of the pre-edge features, while
the main rising edge is not shifted. A similar energy shift
due to the core-hole
was already observed for electric quadrupole transitions
at the Ti $K$ pre-edge in rutile.\cite{Joly99}
Unfortunately, here the shift induced by the presence of
the core-hole is not large enough  to give a perfect agreement
with experiment. This problem may be due to the limit of the
density functional theory within LDA to model the core hole-electron
interaction for empty $d$ states of 3$d$ transition metals.
LDA+U calculations could possibly improve the agreement
in the pre-edge region. Indeed in Ref.\cite{Mazurenko04},
the authors show that
LDA+U calculations performed on ruby essentially
affect the positions of
Cr $d$ level relative to the valence and the conduction band.

\begin{figure*}[hb!]
\includegraphics{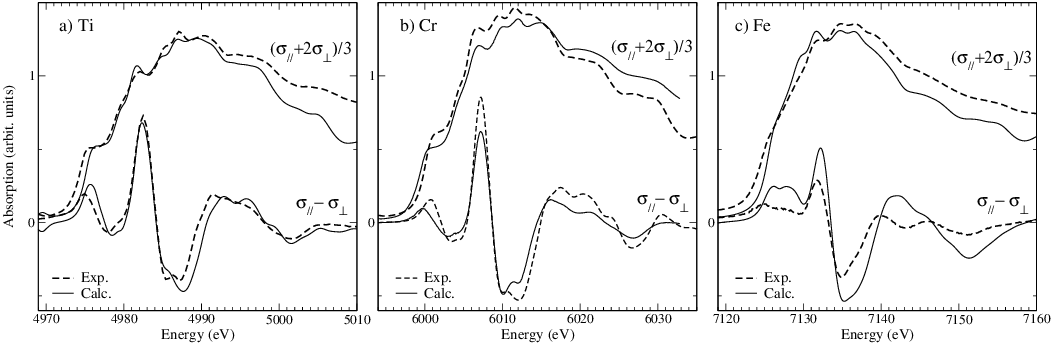}
\caption{\label{exp_calc} Comparison between experimental XANES data
(dashed line) and calculated spectra obtained with the
\textit{relaxed} models (solid line):
a) the Ti $K$ edge in pink sapphire (\pinksapph);
b) the Cr $K$ edge in ruby (\ruby);
c) the Fe $K$ edge in blue sapphire (\bluesapph).
An energy shift was
added to valence band maximum in the calculations
in order to match with experiments:
4966~eV at the Ti $K$edge, 5991~eV at the Cr $K$ edge and
7118~eV at the Fe $K$ edge.}
\end{figure*}

\subsection{\label{subsec:exp_calc} Comparison between experimental 
and calculated XANES spectra} 

Fig.~\ref{exp_calc} compares experimental XANES data with 
the theoretical spectra obtained with the \textit{relaxed} 
models. The $(\sigma_\parallel+2\sigma_\perp)/3$ and 
$\sigma_\parallel-\sigma_\perp$ signals are 
displayed at the Ti $K$ edge for the pink sapphire sample 
(Fig.\ref{exp_calc}a), at the Cr $K$ edge for the ruby 
sample (Fig.\ref{exp_calc}b) and at the Fe $K$ edge for 
the blue sapphire sample (Fig.\ref{exp_calc}c).

One observes that the experimental spectra
present comparable overall shapes. 
The Fe $K$ edge seems to 
be less resolved than the Ti and Cr $K$ edges, probably due
to the following two reasons:
(i) the core-hole lifetime is shorter at the Fe $K$-edge 
(1.25~eV) than at the Cr $K$-edge (1.08~eV) and at
the Ti $K$-edge (0.94~eV);\cite{Krause79} (ii)
the instrumental resolution is estimated to 0.9~eV 
at the Fe $K$-edge while it is 0.6-0.7~eV at both
the Cr and Ti $K$-edges. 
We also noticed similarities 
between the M $K$ edges (M=Ti, Cr, Fe) in
\corund:M$^{3+}$ and in $\alpha$-M$_2$O$_3$ (see Ref.\cite{Grunes83}
for the Cr $K$ edge in \esko\  and the Fe $K$ edge
in \hemat, and Ref.\cite{Waychunas87} for the Ti $K$ 
edge in \tio). At least to a certain extent, these 
spectra could be considered as the signature of a 
transition metal ion with a +III oxydation state in 
six-fold coordinated environment. However it is here 
necessary  to perform calculations to go farther 
in the XANES analysis.

It can be seen in Fig.~\ref{exp_calc} that a good agreement 
is obtained between experimental data and calculated spectra 
carried out on the \textit{relaxed} structural models. 
The agreement is especially good for ruby at the Cr $K$ edge,
where energy positions and relative intensities of the various
features are well reproduced. It should be emphasized that not
only the averaged $(\sigma_\parallel+2\sigma_\perp)/3$
spectra are correctly reproduced, but also
the dichroic signal, from which more precise structural 
information could be extracted. 
Such a good agreement with angular dependent spectra
has been also obtained with the same calculation method: 
at the Al 
$K$ edge in corundum\cite{Cabaret04} and at the Si 
$K$ edge in $\alpha$-quartz.\cite{Taillefumier02}
\begin{figure}[ht!]
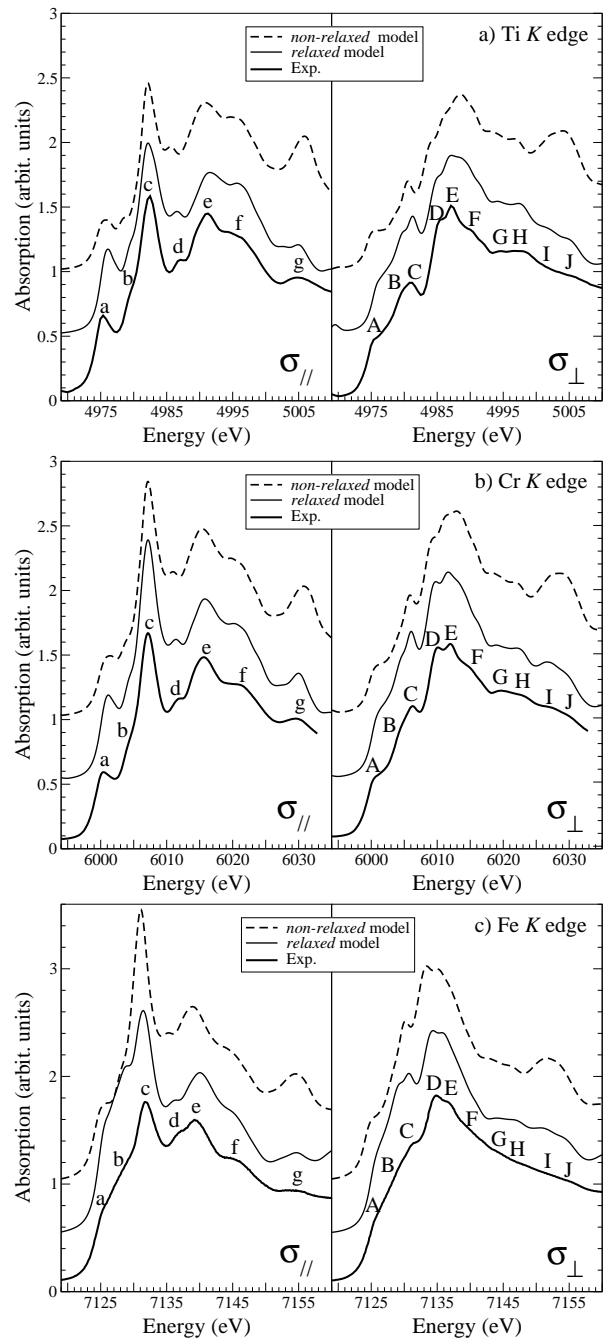

\includegraphics{gaudry_fig7_1.eps}
\includegraphics{gaudry_fig7_2.eps}
\includegraphics{gaudry_fig7_3.eps}
\caption{\label{comp_models} Comparison between
$\sigma_\parallel$ (left) and $\sigma_\perp$ (right)
experimental data (thick solid line) and calculated spectra
obtained with both \textit{relaxed} model (thin solid line)
and \textit{non-relaxed} model (dashed line):
a) the Ti $K$ edge in pink sapphire, \pinksapph; b) the Cr
$K$ edge in ruby, \ruby; c) the Fe $K$ edge in blue sapphire,
\bluesapph.}
\end{figure}

We have seen in Sec.~\ref{subsec:res_models} that the differences
between the \textit{relaxed} and \textit{non-relaxed} structures
are essentially concerned with the interatomic distances within 
the coordination shell
of the impurity. Therefore, it is interesting to test the sensitivity
of XANES to such subtle structural modifications. Fig.~\ref{comp_models}
compares $\sigma_ \parallel$ and $\sigma_\perp$ experimental spectra
with the corresponding calculated ones performed from both
\textit{non-relaxed} and \textit{relaxed} models.
First of all, it is worth noting that the differences observed between 
calculated spectra are small but not negligible.
For all compounds, peaks labeled $g$ in 
$\sigma_\parallel$ and labeled $I$ and $J$ in $\sigma_\perp$
are too strong in  the \textit{non-relaxed} calculated spectra.
Consequently, these features (all located at about 30~eV above the edge) 
are clearly correlated to the first interatomic distances around
the impurity. The same conclusion can be made for the shoulders
labeled $A$ and $B$ in $\sigma_\perp$ and $b$ in $\sigma_\parallel$.
Especially at the Cr $K$ edge in $\sigma_\perp$ signal,
one can see that the double-feature ($D$ and $E$) is better
reproduced by the \textit{relaxed} calculation than by the
\textit{non-relaxed} one.

In a previous study,\cite{Gaudry03} the \textit{relaxed}
model was validated up to the second neighbors around the impurity
by comparison with EXAFS measurements. 
Here, thanks to the very good agreement obtained
between experimental and \textit{relaxed} calculated XANES spectra, 
the \textit{relaxed} structural model is again validated and this time
to greater extent.

\section{Conclusions}

In this paper, we have shown that XANES can provide precious
information about the small structural relaxation occuring within the
atomic site of impurity in an aluminum oxide matrix.
This has been achieved by (i) measuring angular dependent 
XANES spectra in order to probe the distortion of the impurity site,
(ii) calculating structural model by ab initio energy minimization,
(iii) calculating the XANES spectra from the previous 
theoretical structural models using the full-potential 
pseudopotential self-consistent method of Ref.\cite{Taillefumier02}
The good agreement obtained between experimental and calculated
spectra permits the validation of  the
structural models. 
We have also pointed out the importance
of the electronic potential construction to 
carry this study to a successful conclusion.
These results opens new applications in XANES analysis, 
in particular for dilute 
samples for which good quality EXAFS measurements 
are difficult to collect.

\begin{acknowledgments}
We wish to acknowledge the computational support of the French 
\textit{Institut du D\'eveloppement et de Recherche en Informatique 
Scientifique} in Orsay, where all the calculations of this study 
were carried out.
\end{acknowledgments}


\begin{thebibliography}{48}
\bibitem[{E1-()}]{E1-E2}
\bibinfo{note}{For electric dipole and quadrupole transitions, the operator
  $\mathfrak{O}$ of Eq.2 in Ref.~\cite{Taillefumier02} is equal to
  $\hat{\bm{\varepsilon}}\cdot\bm{r}$ and to $\frac{i}{2}\
  \hat{\bm{\varepsilon}}\cdot\bm{r}\ \bm{k}\cdot\bm{r}$, respectively. The unit
  vector $\hat{\bm{\varepsilon}}$ is the direction of the polarisation of the
  incident photon beam, and $\bm{k}$ is the wave vector.}
\bibitem[{qua()}]{quadru}
\bibinfo{note}{One should keep in mind that $(\sigma_\parallel+
  2\sigma_\perp)/3$ is what has been measured. It corresponds to the isotropic
  spectrum only in the electric dipole approximation, but under no
  circumstances in the electric quadrupole approximation.}
\end{thebibliography}
\end{document}